\documentclass[twocolumn,showpacs,prl]{revtex4}

\usepackage{graphicx}
\usepackage{dcolumn}
\usepackage{amsmath}
\usepackage{amssymb}

\makeatletter
\def\btt#1{\texttt{\@backslashchar#1}}%
\DeclareRobustCommand\bblash{\btt{\@backslashchar}}%
\makeatother

\begin{document}

\title[Short Title]{ Luminescence of a Cooper Pair}

\author{
Yasuhiro Asano$^1$, Ikuo Suemune$^{2,3}$, Hideaki Takayanagi$^{3,4,5}$, and Eiichi Hanamura$^6$
}
\affiliation
{
$^1$Department of Applied Physics, Hokkaido University, Sapporo 060-8628, Japan.\\
$^2$Research Institute for Electronic Science, Hokkaido University, Sapporo 001-0021, Japan.\\
$^3$CREST, Japan Science and Technology Agency, Kawaguchi 332-0012, Japan.\\
$^4$Department of Applied Physics, Tokyo University of Science, Tokyo 162-8601, Japan.\\
$^5$International Center for Nanoarichitectonics, NIMS, Tsukuba 305-0044, Japan.\\
$^6$Japan Science and Technology Agency, Kawaguchi 332-0012, Japan. 
}

\date{\today}

\begin{abstract}
This paper theoretically discusses the photon emission spectra of a superconducting pn-junction. 
 On the basis of the second order perturbation theory for electron-photon interaction,
we show that the recombination of a Cooper with two p-type carriers causes drastic 
enhancement of the luminescence intensity. The calculated results of photon emission spectra 
explain characteristic features of observed signal in an recent experiment.
Our results indicate high functionalities of superconducting light-emitting 
devices.
\end{abstract}

\maketitle

Light-emitting diode (LED) usually fabricated on semiconductors has been an important 
element of modern technologies. A trend of research seems to be focusing on 
producing a better controlled 
photon and an entangled photon pair~\cite{michler,benson} for realizing 
quantum computation and quantum information. 
Superconducting devices have an advantage to obtain robustly coherent quantum states 
because of its coherent nature~\cite{chiorescu,wallraff,katz}.
Superconducting LEDs~\cite{hanamura} have been originally proposed in the
context of superradiation. 
They are, however, a promising candidate to create an entangled photon pair~\cite{suemune}. 
A recent theoretical study 
predicts the Josephson radiation in a superconducting pn junction~\cite{recher}. 
Thus superconductor/semiconductor LED hybrids 
undoubtedly have a possibility to produce technologies in the next generation.  

 The radiative recombination of Cooper pairs has been observed recently 
in a InGaAs/InP pn junction attaching onto a superconductor Nb~\cite{hayashi}.
The electroluminescence becomes drastically large at low temperatures 
below the superconducting transition temperature $T_c$ of Nb electrode.
Surprisingly degree of the enhancement in the luminescence intensity is one order of magnitude.
Although the effects of superconductivity on the radiative recombination are clear in 
experiments, a mechanism has been an open problem.
We theoretically address this issue in this paper. 
We study the emission spectra of photon in a superconducting pn 
junction based on the second order perturbation theory.
In the second order expansion, we find that 
a peculiar recombination process to superconductivity enlarges the 
luminescence intensity, where two electrons recombine with two p-type carriers as a Cooper.
The theoretical results explain characteristic features of the experimental 
findings~\cite{hayashi}.
This paper not only figures out a mechanism of the large amplitude of luminescence 
intensity but also gives a guide for designing 
highly functional superconducting light-emitting devices.

\begin{figure}[htbp]
\begin{center}
\includegraphics[width=8.0cm]{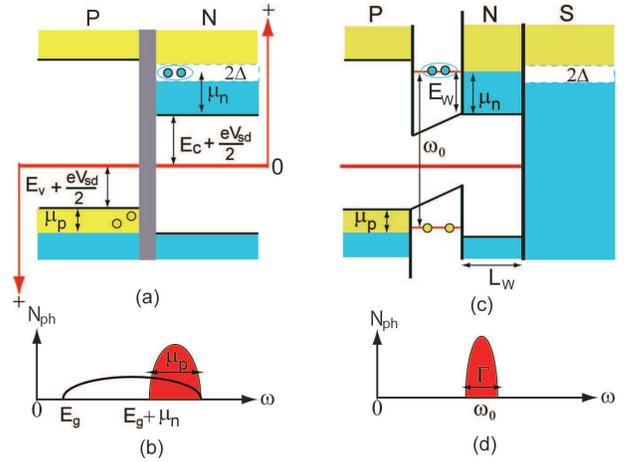}
\end{center}
\caption{
(Color online) 
Schematic energy diagram of pn junctions. 
A theoretical model used for calculation is shown in (a).
In (c), a real junction in an experiment is illustrated. 
Theoretical results of the photon spectra are shown in (b) and (d). 
}
\label{fig1}
\end{figure}
Let us consider a p-type semiconductor / superconductor junction as shown in Fig.~\ref{fig1}(a). 
The energy is measured from the horizontal line indicated by '0'.
The sign of energy in a p-type semiconductor is chosen to be opposite to that in a superconductor. 
We assume that a semiconductor and a superconductor are in their local equilibrium which 
are characterized by the local chemical potential $\mu_p$ and $\mu_n$, respectively. 
The edges of the conduction and valence bands are $E_c$ and $E_v$, respectively. 
In what follows, we use a unit of $\hbar=k_B=c=1$, where $k_B$ is the Boltzmann constant 
and $c$ is the speed of light.
The p-type semiconductor is described by
\begin{align}
H_p=\sum_{\boldsymbol{k},\sigma} \epsilon_p(k) b_{\boldsymbol{k},\sigma}^\dagger 
b_{\boldsymbol{k},\sigma}^{ },\label{hp}
\end{align}
where $\epsilon_p(k)= {k}^2/(2m_p) +E_v + eV_{sd}/2$, 
$m_p$ is the effective mass, $V_{sd}$ is the applied bias voltage across the junction, and 
$b_{\boldsymbol{k},\sigma}^\dagger (b_{\boldsymbol{k},\sigma}^{ })$ is the creation 
(annihilation) operator of a p-type carrier with a wave 
number $\boldsymbol{k}$ and spin $\sigma= \uparrow$ or $\downarrow$.
The photon states is described by 
\begin{align}
H_{ph}=&\sum_{\boldsymbol{q}} \omega_q  
\left(a_{\boldsymbol{q}}^\dagger a_{\boldsymbol{q}}^{ } +\frac{1}{2}\right), 
\end{align}
where $a_{\boldsymbol{q}}^\dagger (a_{\boldsymbol{q}}^{ })$ is the creation (annihilation)
operator of a photon with a wave number $\boldsymbol{q}$ and an energy $\omega_q$.
The normal state in a metal is described by
 \begin{align}
H_{nn}=& \sum_{\boldsymbol{k},\sigma} \left( \frac{k^2}{2m_n} + E_c+ \frac{eV_{sd}}{2}\right) 
c_{\boldsymbol{k},\sigma}^\dagger c_{\boldsymbol{k},\sigma}^{ },
\end{align}
with $m_n$ being the effective mass. 
The electron-photon interaction Hamiltonian reads 
\begin{align}
H_I=\sum_{\boldsymbol{k},\boldsymbol{q},\sigma} B_{\boldsymbol{k},\boldsymbol{q}}\,
b_{\boldsymbol{k}-\boldsymbol{q},\sigma}\, 
\, c_{\boldsymbol{k},\sigma}\, a_{\boldsymbol{q}}^\dagger + \text{H.c.},
\end{align}
where $B_{\boldsymbol{k},\boldsymbol{q}}$ is the coupling energy.
On the basis of the second order 
perturbation theory, 
the number of photon $N_{ph}=\sum_{\boldsymbol{q}}a_{\boldsymbol{q}}^\dagger 
a_{\boldsymbol{q}}^{ }$
is calculated as
\begin{align}
&\langle N_{ph} \rangle = \langle N_{ph}(1) \rangle + \langle N_{ph}(2) \rangle,\\
&\langle N_{ph}(1) \rangle= \int_{-\infty}^t \!\!\!\!\! dt_1\int_{-\infty}^t \!\!\!\!\!
 dt_2\langle \chi_0 | H_I(t_1) 
N_{ph}
 H_I(t_2)|\chi_0 \rangle,\\
&\langle N_{ph}(2) \rangle=\int_{-\infty}^{t_2} \!\!\!\!\! dt_1\int_{-\infty}^t \!\!\!\!\! dt_2
\int_{-\infty}^{t} \!\!\!\!\! dt_3\int_{-\infty}^{t_3} \!\!\!\!\! dt_4 \; I(2),\\
&I(2)= \langle \chi_0 |  H_I(t_1)  H_I(t_2)  N_{ph}  H_I(t_3)  H_I(t_4)
|\chi_0 \rangle,\\
&|\chi_0 \rangle \to |0\rangle \otimes |N\rangle \otimes |P\rangle, \label{chi0}
\end{align}
where $|0\rangle$ is the zero photon state.

The BCS theory describes superconducting states,
\begin{align}
H_{ns}=& \sum_{\boldsymbol{k},\sigma} E_k  
\gamma_{\boldsymbol{k},\sigma}^\dagger \gamma_{\boldsymbol{k},\sigma}^{ }, \label{hns}
\end{align}
where $E_k= \sqrt{ \xi_n^2(k)  + \Delta^2 }$, $\xi_n(k)={k^2}/{2m_n}-\mu_n$, 
$\Delta$ is the pair potential, and
$\gamma_{\boldsymbol{k},\sigma}^\dagger (\gamma_{\boldsymbol{k},\sigma}^{ })$ is 
the creation (annihilation) operator of Bogoliubov quasiparticle. 
 This description, however, is valid within a small energy scale near the Fermi level
which is at $\tilde{\mu}_n= E_c + eV_{sd}/2+\mu_n$ measured from '0'.  
To apply the BCS theory to the present issue, a rule is necessary to describe 
the operator in the interaction picture. 
The canonical transformation connects an electron operator and Bogoliubov operators by
\begin{align}
c^\dagger_{\boldsymbol{k},\sigma}(t)=e^{i\tilde{\mu}_n t} \left( u_k e^{iE_kt} \gamma_{\boldsymbol{k},\sigma}^\dagger\
-s_{\sigma} v_ke^{-iE_kt}\gamma_{-\boldsymbol{k},\bar{\sigma}}^{ }\right),
\end{align}
in $\langle \chi_0| \cdots |\chi_0\rangle$, where $u_k(v_k)=[(1+(-)\xi_n(k)/E_k)/2]^{1/2}$, 
$s_{\sigma}=1 (-1)$ for $\sigma=\uparrow (\downarrow)$, and $\bar{\sigma}$ means 
the opposite spin to $\sigma$.
The thermal average of operators is carried out in the local equilibrium.
In a p-type semiconductor, for instance, the average of operators are calculated in 
\begin{align}
H_p'=\sum_{\boldsymbol{k},\sigma} \xi_p(k) b_{\boldsymbol{k},\sigma}^\dagger 
b_{\boldsymbol{k},\sigma}^{ },
\end{align}
instead of Eq.~(\ref{hp}) with $\xi_p(k)=k^2/2m_p-\mu_p$. 
In a superconductor, the average of the Bogoliubov operators
are calculated in Eq.~(\ref{hns}). 
In Eq.~(\ref{chi0}), $|P\rangle$ means the state vector of 
p-type carrier in the local equilibrium and $|N\rangle$  
indicates the BCS state in the local equilibrium.

The time average of the photon number $\overline{\langle N_{ph}\rangle}$ corresponds
 to the luminescence intensity and it in the first order perturbation expansion results in
\begin{align}
&\overline{\langle N_{ph}(1) \rangle } = 2\pi \sum_{\boldsymbol{k}, \boldsymbol{q}, \sigma} |B_{\boldsymbol{k},\boldsymbol{q}}|^2 f^{p}_{\boldsymbol{k}-\boldsymbol{q}} \nonumber\\
&\times\left[
u_k^2 f^n_k \delta( \tilde{\omega} - E_k ) + v_k^2(1-f^n_k) \delta( \tilde{\omega} + E_k ) \right],
\label{n1}
\end{align}
where $\tilde{\omega}=\omega_q-E_g - \mu_n - \mu_p - \xi_p(\boldsymbol{k}-\boldsymbol{q})$, $E_g = E_v + E_c + eV_{sd}$, $f^{n}_{k}=[1- \tanh(E_k/2T)]/2$, and
$f^{p}_{k}=[1- \tanh(\xi_p(k)/2T)]/2$.
This result recovers the photon spectra in a normal pn junction by tuning $\Delta\to 0$, which 
means that $E_k \to - \xi_n(k), u_k\to 0$, $v_k \to 1$ for $k<k_F$ and 
$E_k \to \xi_n(k), u_k\to 1$, $v_k \to 0$ for $k> k_F$ with $k_F$ being 
the Fermi wave number satisfying $k_F^2/2m_n=\mu_n$.
The threshold of spectra is $E_g$ and the width of spectra is given by $\mu_n+\mu_p$.  
The spectra in Eq.~(\ref{n1}) have a broad profile reflecting the quasiparticle 
density of states as shown in Fig.~\ref{fig1}(b). 

One of the characteristics peculiar to superconductivity is a
giant oscillator strength~\cite{hanamura}. 
This comes from the much freedom of the wave vector $\boldsymbol{k}$ 
for the remaining elementary excitation when a 
Cooper pair decays radiatively with a p-type carrier with the 
opposite wave vector $-\boldsymbol{k}$. 
This would gives the large luminescence intensity in the experiment~\cite{hayashi}.
 This effect of the giant oscillator strength works also in the first 
stage in the second order expansion while not in the second
process. The other characteristics is coming from the resonant enhancement
due to the nearly degenerate nature of the final and 
intermediate states close to the initial state. This is shown in the following 
calculation.

\begin{figure}[htbp]
\begin{center}
\includegraphics[width=8.0cm]{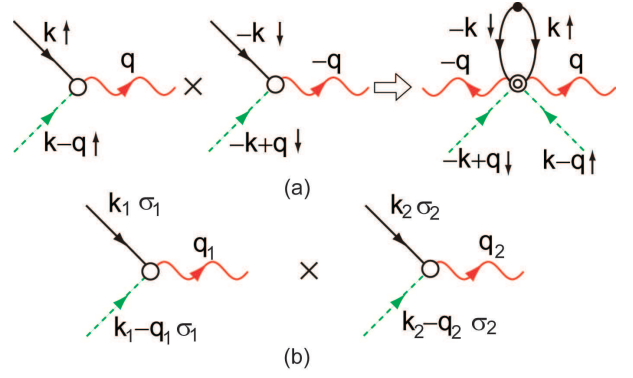}
\end{center}
\caption{(Color online)
Recombination processes in the second order perturbation expansion, where
solid, broken and wavy lines represent the propagation of an electron, 
a p-type carrier and a photon, respectively 
In (a), a recombination of a Cooper pair in $Q_N(S)$ is shown.
In (b), a recombination process other than $Q_N(S)$ is illustrated.
}
\label{fig2}
\end{figure}

The results of the second order perturbation are given by
\begin{align}
I(2)=&\!\!\!\!\!\!
\sum_{\boldsymbol{k}_1\cdots \boldsymbol{k}_4, \boldsymbol{q}_1 \cdots \boldsymbol{q}_4, \sigma_1 \cdots \sigma_4} 
\!\!\!\!\!\!\!\!\!\!\!\!\!\!\!\!\!\!
e^{-i\Omega_1t_1-i\Omega_2t_2 +i\Omega_3 t_3 +i \Omega_4 t_4}\nonumber\\
&\times B^\ast_{\boldsymbol{k}_1,\boldsymbol{q}_1} B^\ast_{\boldsymbol{k}_2,\boldsymbol{q}_2}
B_{\boldsymbol{k}_3,\boldsymbol{q}_3}B_{\boldsymbol{k}_4,\boldsymbol{q}_4}
Q_{Ph} Q_P Q_N,\label{i2}
\end{align}
with 
$\Omega_j(\boldsymbol{k}_j,\boldsymbol{q}_j)= \omega_{q_j} - \epsilon_P(\boldsymbol{k}_j-\boldsymbol{q}_j)-\tilde{\mu}_n$. The average of the operators
$Q_{Ph}$, $Q_P$ and $Q_N$ are calculated as follows,
\begin{align}
Q_{Ph}=&\sum_{\boldsymbol{q}_5}
\langle0| a_{\boldsymbol{q}_1}^{ }a_{\boldsymbol{q}_2}^{ }a_{\boldsymbol{q}_5}^\dagger a_{\boldsymbol{q}_5}^{ }a_{\boldsymbol{q}_3}^\dagger a_{\boldsymbol{q}_4}^\dagger
|0\rangle,  
\nonumber \\
=&2( \delta^{\boldsymbol{q}}_{14} \delta^{\boldsymbol{q}}_{23}+\delta^{\boldsymbol{q}}_{13}
\delta^{\boldsymbol{q}}_{24}),\label{qph}\\
Q_P=& \langle P| b^\dagger_{\boldsymbol{p}_1,\sigma_1} b^\dagger_{\boldsymbol{p}_2,\sigma_2}
b^{ }_{\boldsymbol{p}_3,\sigma_3}b^{ }_{\boldsymbol{p}_4,\sigma_4} |P\rangle,\nonumber \\
=& f_{\boldsymbol{p}_1}^p f_{\boldsymbol{p}_2}^p
(\delta^{\sigma}_{14} \delta^{\sigma}_{23}
\delta^{\boldsymbol{p}}_{14}\delta^{\boldsymbol{p}}_{23}
-\delta^{\sigma}_{13}\delta^{\sigma}_{24}
\delta^{\boldsymbol{p}}_{13}\delta^{\boldsymbol{p}}_{24}),\label{qp}
\end{align}
where $\delta^{\boldsymbol{q}}_{ij}=\delta_{\boldsymbol{q}_i,\boldsymbol{q}_j}$, 
$\delta^{\sigma}_{ij}=\delta_{\sigma_i,\sigma_j}$, 
$\delta^{\boldsymbol{p}}_{ij}=\delta_{\boldsymbol{p}_i,\boldsymbol{p}_j}$ and 
$\boldsymbol{p}_j= \boldsymbol{k}_j-\boldsymbol{q}_j$. By applying the Bogoliubov transformation, 
 we find,
\begin{align}
&Q_N= \langle N|
(u_{k_1} e^{iE_{k_1}t_1} \gamma_{\boldsymbol{k}_1,\sigma_1}^\dagger - \sigma_1 v_{k_1} 
e^{-iE_{k_1}t_1} \gamma_{-\boldsymbol{k}_1,\bar{\sigma}_1}^{ })  \nonumber\\
& \times
(u_{k_2}e^{iE_{k_2}t_2}\gamma_{\boldsymbol{k}_2,\sigma_2}^\dagger - \sigma_2 v_{k_2} 
e^{-iE_{k_2}t_2}\gamma_{-\boldsymbol{k}_2,\bar{\sigma}_2}^{ }) \nonumber\\
& \times
(u_{k_3}e^{-iE_{k_3}t_3}\gamma_{\boldsymbol{k}_3,\sigma_3}^{ } - \sigma_3 v_{k_3} 
e^{iE_{k_3}t_3} \gamma_{-\boldsymbol{k}_3,\bar{\sigma}_3}^\dagger) \nonumber\\
& \times 
(u_{k_4}e^{-iE_{k_4}t_4}\gamma_{\boldsymbol{k}_4,\sigma_4}^{ } - \sigma_4 v_{k_4} 
e^{iE_{k_4}t_4} \gamma_{-\boldsymbol{k}_4,\bar{\sigma}_4}^\dagger) 
|N\rangle, \label{qn}
 \end{align}
which gives twelve terms.
In what follows, we extract the most dominant contribution in the second order terms.
The average of $Q_N$ includes following four terms 
\begin{align}
Q_N&(S)=u_{k_1}v_{k_1}u_{k_3}v_{k_3} \delta_{\sigma_1,\bar{\sigma}_2} 
\delta_{\sigma_3,\bar{\sigma}_4} \delta_{\boldsymbol{k}_1,-\boldsymbol{k}_2} \delta_{\boldsymbol{k}_3,-\boldsymbol{k}_4} \sigma_1\sigma_3
\nonumber\\
&\times \left[
  e^{iE_{k_1}(t_1-t_2)} e^{-iE_{k_3}(t_3-t_4)} f^n_{k_1}(1-f^n_{k_3}) \right. 
\nonumber\\
&+   e^{-iE_{k_1}(t_1-t_2)} e^{iE_{k_3}(t_3-t_4)} f^n_{k_3}(1-f^n_{k_1}) 
\nonumber\\
&- e^{-iE_{k_1}(t_1-t_2)} e^{-iE_{k_3}(t_3-t_4)} 
(1-f^n_{k_1})(1-f^n_{k_3})
\nonumber\\
&\left.-   e^{iE_{k_1}(t_1-t_2)} e^{iE_{k_3}(t_3-t_4)} f^n_{k_1}f^n_{k_3}
  \right].\label{qn1}
\end{align}
Because 
$\delta_{\sigma_3,\bar{\sigma}_4} \delta_{\boldsymbol{k}_3,-\boldsymbol{k}_4}$ in Eq.~(\ref{qn1})
means the destruction of two electrons as a Cooper pair, 
$Q_N(S)$ describe effects of superconductivity on the emission spectra.
Another eight terms in $Q_N$ contribute to the emitting processes
described by Fig.~\ref{fig2}(b) which gives the luminescence intensity
proportional to $\overline{\langle N_{ph}(1)} \rangle^2$. 
We will show that $Q_N(S)$ gives a large contribution to the 
emission spectra at $\delta\boldsymbol{q}=\boldsymbol{q}_1+\boldsymbol{q}_2=0$, 
($\delta\boldsymbol{q}=\boldsymbol{q}_3+\boldsymbol{q}_4=0$ in other wards)~\cite{smallq}.
Substituting Eqs.~(\ref{qph}),(\ref{qp}) and (\ref{qn1}) into Eq.~(\ref{i2}) 
and carrying out time integrations,  
we obtain 
\begin{align}
\overline{\langle N_{ph}(2) \rangle } =&   4\pi |B|^4\sum_{\boldsymbol{q},\sigma} 
\delta(\Omega_{\boldsymbol{k}_F ,\boldsymbol{q}})I_0, \label{n2-1}\\
I_0=& \sum_{\boldsymbol{k}} \left[   \frac{f^n_{k}(1-f^n_{k})}{(E_k-i/\tau)^2} + \frac{f^n_{k}(1-f^n_{k})}{(E_k+i/\tau)^2}
  \right.\nonumber \\
& \left.+ \frac{( f^n_{k})^2 + (1- f^n_{k})^2}{E_k^2+(1/\tau)^2}
\right] \frac{\Delta^2}{E_{k}^2},\label{i0}
\end{align}
where we introduce a relaxation time $\tau$ to 
remove effects of the perturbation at $t \to -\infty$, 
we neglect dependence of $B$ on wave numbers and assume $f^p_{\boldsymbol{k}_F-\boldsymbol{q}}=1$.
At $1/\tau=0$, $I_0=\pi N_0 / 2 \Delta$ essentially 
diverges for small $\Delta$ with $N_0$ denoting the normal density of states 
in a superconductor at the Fermi energy. 
The singular behavior at small $\Delta$ in Eq.~(\ref{i0})
is a sign of the large luminescence intensity due to superconductivity.
%
%

We first show mathematical reasons of the singularity. Then we will discuss
 the physics behind the phenomenon. 
A two-photon emitting process in $Q_N(S)$ 
is illustrated in Fig.~\ref{fig2}(a).
The annihilation of a Cooper pair is described by 
$c_{-\boldsymbol{k},\downarrow}c_{\boldsymbol{k},\uparrow}$
which includes a operator 
$\gamma_{\boldsymbol{k},\uparrow}^\dagger\gamma_{\boldsymbol{k},\uparrow}^{ }$.
Let us assume that the energy of the initial state is zero.
In the first order expansion, the operation of $\gamma_{\boldsymbol{k},\uparrow}$ to the BSC state 
decreases energy by $E_k+\tilde{\mu}_n$. At the same time,
a p-type carrier with energy $\epsilon_p(\boldsymbol{k}-\boldsymbol{q})$ is destructed 
and a photon with energy $\omega_q$ is created.
Thus the energy of the intermediate state $\delta E_1$ results in 
$\delta E_1=\omega_q - \epsilon_p(\boldsymbol{k}-\boldsymbol{q}) - E_k - \tilde{\mu}_n=\Omega_{\boldsymbol{k},\boldsymbol{q}}-E_k$ which
is the energy denominator in the perturbation expansion. 
In the second order, the operation of $\gamma_{\boldsymbol{k},\uparrow}^\dagger$,
 the destruction of a p-type carrier, 
and the creation of a photon gain energy by 
$E_k - \tilde{\mu}_n$, $-\epsilon_p(-\boldsymbol{k}+\boldsymbol{q})$, and $\omega_{-q}$, 
respectively. 
Therefore the difference in energy between the intermediate state and the final one
becomes $\delta E_2= \omega_q -\epsilon_p(\boldsymbol{k}-\boldsymbol{q})+ E_k - \tilde{\mu}_n
=\Omega_{\boldsymbol{k},\boldsymbol{q}}+E_k$. 
The perturbation theory requires the energy conservation between 
the initial and the final states, (i.e., $\delta E_1 +\delta E_2=0$), which 
leads to $2\Omega_{\boldsymbol{k},\boldsymbol{q}}=0$.
As a result, only a small value of $E_k$ remains
in the denominator as shown in Eq.~(\ref{i0}).
The physics behind the phenomena is simple. 
The BCS state can have ability to emit a pair of photons with 
remaining its state almost unchanged because 
the BCS state is the eigen state of
$\gamma_{\boldsymbol{k},\uparrow}^\dagger\gamma_{\boldsymbol{k},\uparrow}^{ }$.
The equation $\Omega_{\boldsymbol{k}_F,\boldsymbol{q}}=0$ is the condition for emitting a photon.
The threshold and width of spectra are $E_g+\mu_n$ and $\mu_p$, respectively.
In Fig.~\ref{fig1}(b), we show predicted spectra in the second order process. 

\begin{figure}[htbp]
\begin{center}
\includegraphics[width=8.0cm]{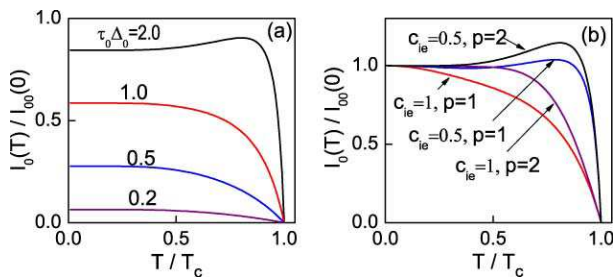}
\end{center}
\caption{
Temperature dependence of luminescence intensity.
In (a), we consider relaxation time due to the elastic impurity scatterings 
by $\alpha=\tau_{0}\Delta_0$. 
In (b), we consider the relaxation due to inelastic scatterings.
}
\label{fig3}
\end{figure}

The singular behavior in perturbation expansion 
implies an importance of higher order terms to predict the luminescence 
intensity quantitatively. 
Here we do not discuss this issue, but choose an alternative way 
of regularizing the obtained results for qualitative argument. 
In what follows, we introduce a finite relaxation time.
First we consider mean free time due to elastic impurity
scatterings $\tau_{0}$.
At $T=0$, we obtain $I_0 = I_{00}(0) 2\alpha^2/( \sqrt{1+\alpha^2}(\alpha + \sqrt{1+\alpha^2}))$,
where $\alpha=\tau_{0}\Delta_0$, $\Delta_0$ is the pair potential at the zero temperature and 
$I_{00}(0)=\pi N_0 / 2 \Delta_0$ is Eq.~(\ref{i0}) at $T=0$. 
At $T \lesssim T_c$, we find 
\begin{align}
\frac{I_0}{I_{00}(0)} \approx \left\{
\begin{array}{ll} c_0 \alpha^2 (\Delta/\Delta_0)^2 \Delta_0/ T & \alpha\lesssim 1\\  
\alpha^3 (\Delta/\Delta_0)^2 & \alpha \gg 1,
\end{array}\right.
\end{align}
where $c_0$ is a constant of the order of unity.
In Fig.~\ref{fig3}(a), we show $I_0$ as a function of temperature for several choices of 
$\alpha$, where we describe the dependence of $\Delta$ on temperature by the BCS theory.
The amplitude of $I_0$ at $T=0$ is suppressed in the dirty limit
as shown in a result with $\alpha=0.2$. 
The amplitude at $T=0$ increases with increasing $\alpha$.
At $\alpha=1$, $I_0(0)$ has almost the same amplitude as
$I_{00}(0)$. When we increase $\alpha$ up to 2.0, the results show a bump just below $T_c$.
Next we consider inelastic scatterings described by 
$1/{\tau_{ie}}= C_{ie} \left( {T}/{T_c} \right)^p$, 
where $C_{ie}$ is a coupling constant and $p$ depends on 
scattering sources such as $p=1$ for electron-phonon scatterings and $p=2$ 
for repulsive electron-electron interaction.
In Fig.~\ref{fig3}(b), we calculate $I_0$ for several choices of 
$C_{ie}$ and $p$. Since $1/\tau_{ie} \to 0 $ at $T=0$, the amplitude 
is close to $I_{00}(0)$ at $T=0$. When we decreases $C_{ie}$, the bump 
appears below $T_c$ as well as in (a). 
For $1/\tau \lesssim \Delta_0$,  
the luminescence intensity at $T \lesssim T_c$ is then given by
\begin{align}
\overline{\langle N_{ph}(2) \rangle }& \approx 4 \pi c_0 |B|^4 N_0
 \frac{ (\tau \Delta)^2}{ T} \sum_{q}\delta(\Omega_{k_F,q}).
\label{n2}
\end{align}

Finally we modify Eq.~(\ref{n2}) to describe the photon spectra in the 
experiment~\cite{hayashi} as shown in Fig.~\ref{fig1}(c).
In the real junction, a superconductor is attached to a n-type semiconductor
whose thickness is $L_w$.
The proximity effect enhances the luminescence intensity.
In Eq.~(\ref{n2}), $\Delta$ is proportional to the amplitude of a Cooper pair.
In n-type semiconductor, the proximity effect enables the pair amplitude
which proportional to $\Delta e^{-L_w/\xi_T}$ with $\xi_T=\sqrt{D/2\pi T}$ 
and $D$ being the diffusion constant in the n-type semiconductor. 
The photon pairs are emitted in a quantum well between the p- and n-type semiconductors.
The level in the quantum well $E_w$ should coincide with the Fermi level
in the n-type semiconductor $\mu_n$. Namely $|E_w-\mu_n|$ must be less than 
both the Thouless energy $E_{Th}=D/L_w^2$ and $\Delta$.
This resonant condition is particularly important for a Cooper pair to penetrate
into the quantum well. The emission spectra has a peak at 
$\omega_0$ and the peak width is given by $\Gamma=t_w^2 N_0$, where $t_w$ is the 
transfer integral between the quantum well and the semiconductor. 
The argument above is summarized by an equation for $T \lesssim T_c$
\begin{align}
\overline{\langle N_{ph}(2) \rangle } \approx  |B|^4N_0 \Gamma \sum_{\boldsymbol{q},\sigma}  
\frac{\Delta^2 \tau^2 e^{-2L_W/\xi_T}/ T}{(\omega_q - \omega_0)^2 + (\Gamma)^2},
\label{final}
\end{align}
where we introduce the Lorentz resonant function by hand.
In the experiment, $\xi_T$ is estimated to be much larger than $L_w$ below $T_c$.
Thus the theoretical results in Fig.~\ref{fig3} 
can describe experimental results of the luminescence intensity.
In fact, the experimental results 
of Fig.~6(b) in Ref.~\onlinecite{hayashi} show a very similar line shape 
to that in Fig.~\ref{fig3}(a) with $\alpha=1$. 
%

In conclusion, we have studied the photon emission spectra in a superconducting 
pn junction based on the second order perturbation theory for electron-photon interaction. 
We have found in the second order expansion that 
a peculiar recombination process to superconductivity enlarges the 
luminescence intensity.  The theoretical results explain temperature dependence of 
the luminescence intensity observed in an recent experiment.


{} 


\begin{thebibliography}{}
\bibitem{smallq} The speed of light is much larger than the Fermi velocity. 
In such case, this condition becomes $|\boldsymbol{q}_1|\sim |\boldsymbol{q}_2|$.
\end{thebibliography}
\end{document}